\documentclass[aps,twocolumn,showpacs,preprintnumbers]{revtex4-1}

\usepackage{graphicx}
\usepackage{dcolumn}
\usepackage{bm}
\usepackage{booktabs}

\begin{document}

\preprint{APS/123-QED}

\title{Filamentary superconductivity in semiconducting policrystalline  ZrSe$_2$ compound with Zr vacancies.}
\author{Gabriela Tenorio, L. Bucio${^1}$, and R Escudero*}.
\address{Instituto de Investigaciones en Materiales, Universidad Nacional Aut\'{o}noma de M\'{e}xico,
Apartado Postal 70-360,  M\'{e}xico D.F. 04510, M\'{E}XICO}
\address{$^1$Instituto de F\'{i}sica, Universidad Nacional Aut\'{o}noma de M\'{e}xico,  M\'{e}xico D.F. 04510, M\'{E}XICO}

\email[Author to whom correspondence should be addressed. RE, email address:] {escu@unam.mx}

\date{\today}

\begin{abstract}

$ZrSe_2$ is a band semiconductor studied long time ago. It has  interesting  electronic properties, and because its layers structure can be intercalated with different atoms to change some of the physical properties. In this investigation   we found that Zr deficiencies alter the semiconducting behavior and the compound can be  turned into a superconductor. In this paper  we report our studies related to this discovery. The decreasing of the number of  Zr atoms in small proportion according to the formula Zr$_x$Se$_2$, where  $x$ is varied from  about 8.1 to  8.6 K,  changing the semiconducting behavior to a  superconductor with   transition temperatures ranging between 7.8 to  8.5 K, it   depending of the deficiencies. Outside of those ranges   the compound behaves as  semiconducting with the  properties already known. In our experiments we found that this  new superconductor has only a very small fraction of  superconducting material determined  by   magnetic measurements with applied magnetic field of 10 Oe. Our  conclusions is that superconductivity is  filamentary. However, in one  studied sample  the fraction  was  about 10.2 \%, whereas in others is  only about 1 \% or less.  We determined the superconducting characteristics;  the critical fields  that indicate    a type two superonductor  with Ginzburg-Landau  $\kappa$ parameter of the order about 2.7. The synthesis procedure  is quite normal following the  conventional solid state reaction. In this paper are  included, the electronic characteristics,   transition temperature, and evolution with temperature of the critical fields.

\end{abstract}

\maketitle

\section{Introduction}

\begin{figure}
\begin{center}
\includegraphics[scale=.32]{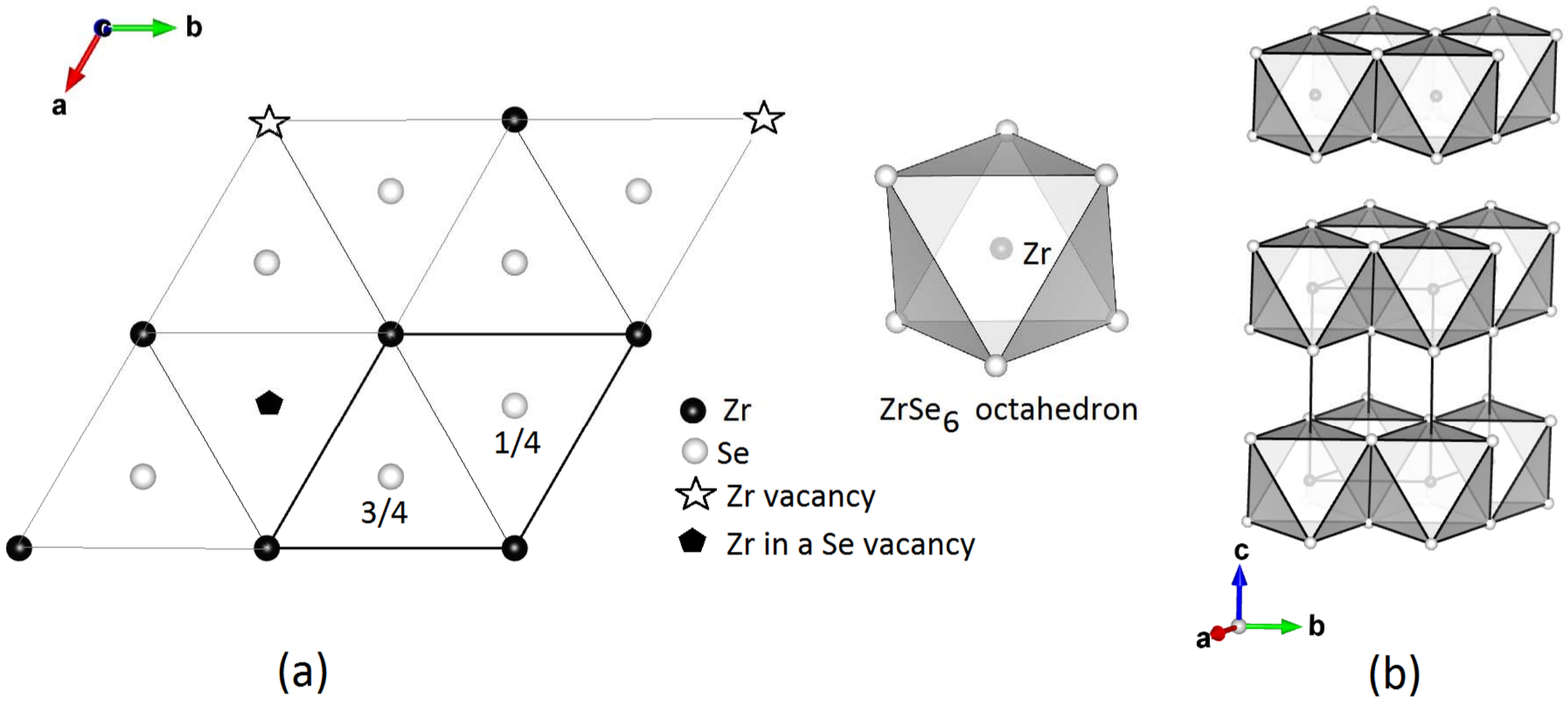}
\caption{(Color on-line) Crystalline structure of ZrSe$_2$ and deficiencies. In a) we  show Zr in Se positions b) shows the stacking structure of the compound, in the middle is shown the ZrSe$_6$ octhahedron. This stacking  structure permits the intercalation of small atoms. }
\end{center}
\end{figure}

\begin{figure}
\begin{center}
\includegraphics[scale=.4]{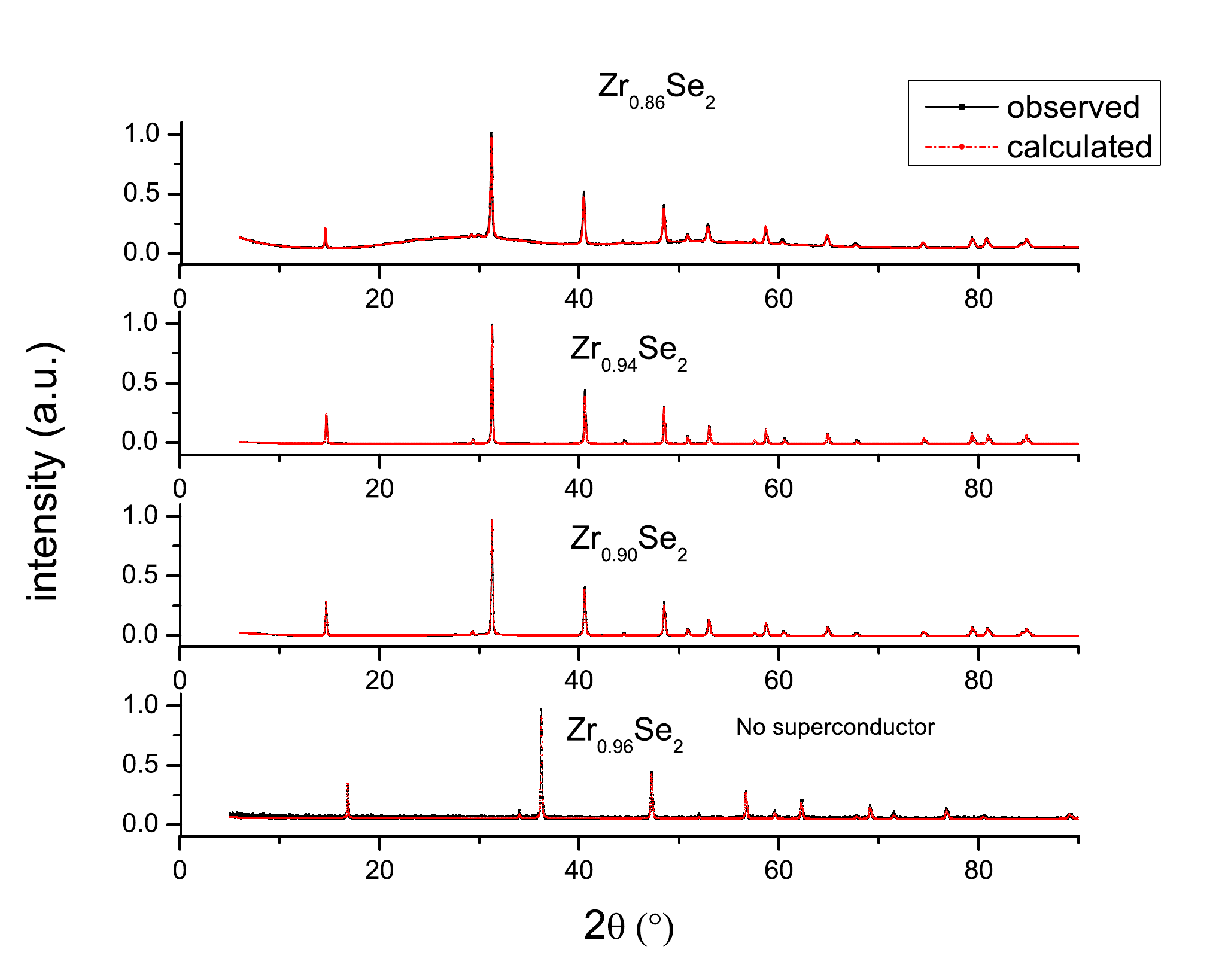}
\caption{(Color on-line) Difractograms of four samples and compositions,   the bottom  difractogram with maximum Zr content  about 0.96\% was not superconducting. The three next difractograms with Zr content 0.90, 0.94, and 0.86 were superconducting. The crystalline structure was Rietveld  refined and shows complete agreements to ZrSe$_2$ close to the stoichiometric composition, without impurities.  Slightly above 0.94. the compound is  semiconducting. The small black dots in the figures are the observed data, and the red lines are the Riettveld fitting. }
\end{center}
\end{figure}

\begin{figure}
\begin{center}
\includegraphics[scale=.4]{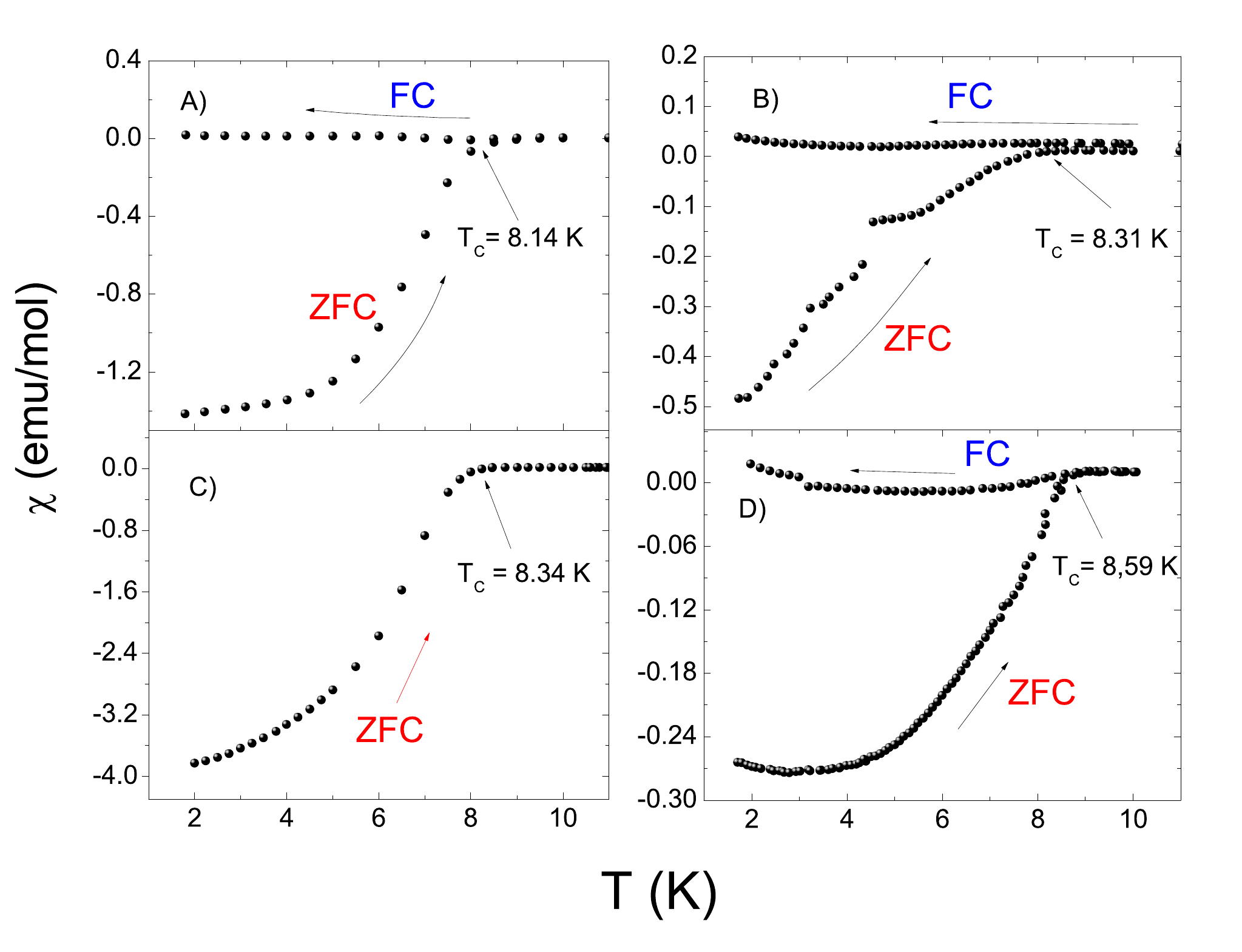}
\caption{(Color on-line) Superconducting transition temperature determined by magnetic measurements with field of 10 Oe in two modes of measurements; Zero Field Cooling (ZFC) and Field Cooling (FC). The curves show small differences in transition temperature but appreciables.  Arrows mark the onset temperature from 8.14 K, 8.31 K, 8.34 K and 8.59 K. The fraction of superconducting material was very small, only in Fig C the amount was about 10.2 \%, other were   about 1 \% or small. }
\end{center}
\end{figure}

\begin{figure}
\begin{center}
\includegraphics[scale=.4]{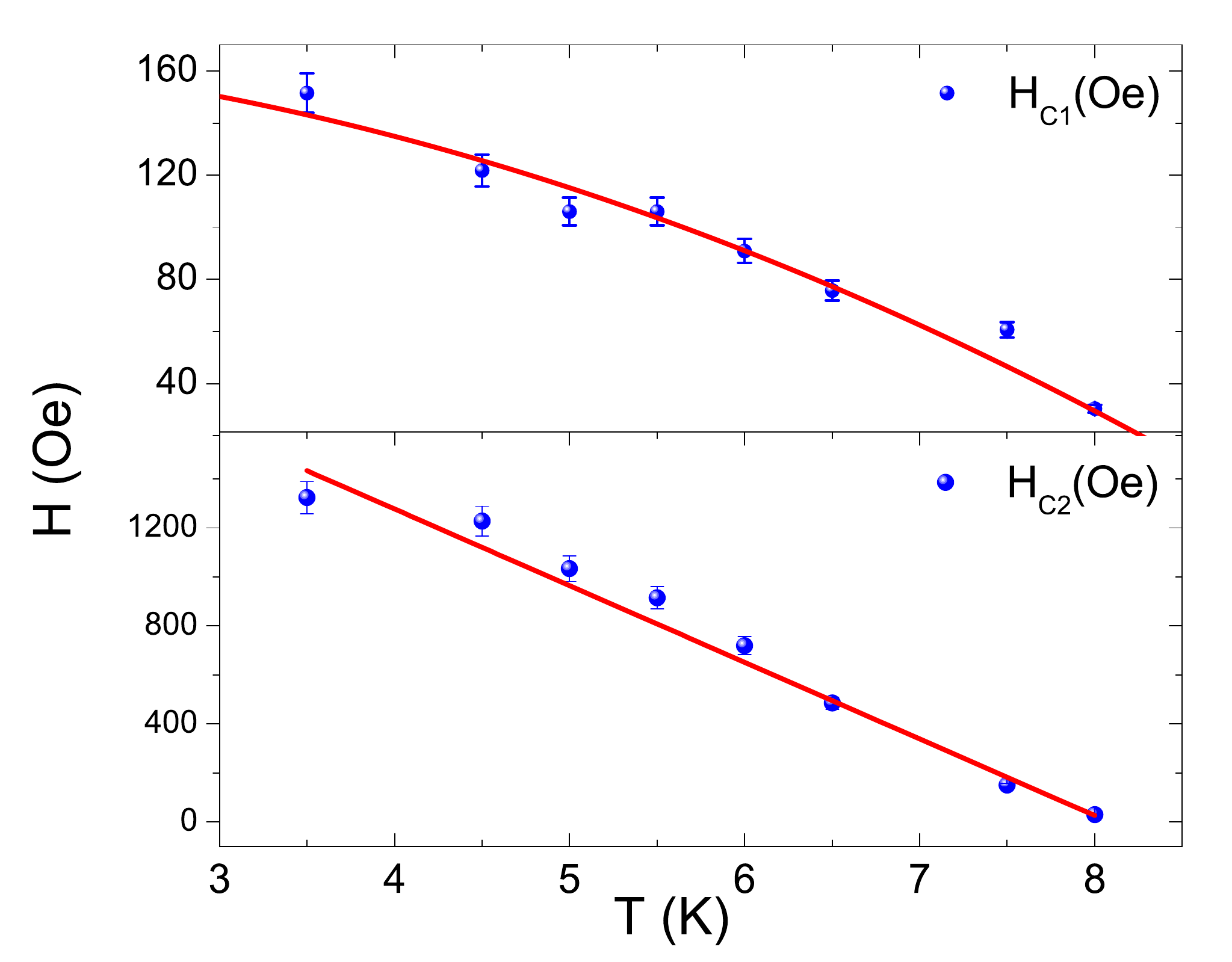}
\caption{(Color on-line) Critical magnetic fields H$_{C1}$, and H$_{C2}$ versus temperature. H$_{C1}$ at the lower temperature is about 170 Oe, whereas H$_{C2}$ is about 2400 Oe. H$_{C1}$ fits quite well to a parabolic behavior with transition temperature at 8.8 K, but the best fit for H$_{C2}$ was a  linear function. }
\end{center}
\end{figure}

\begin{figure}
\begin{center}
\includegraphics[scale=.36]{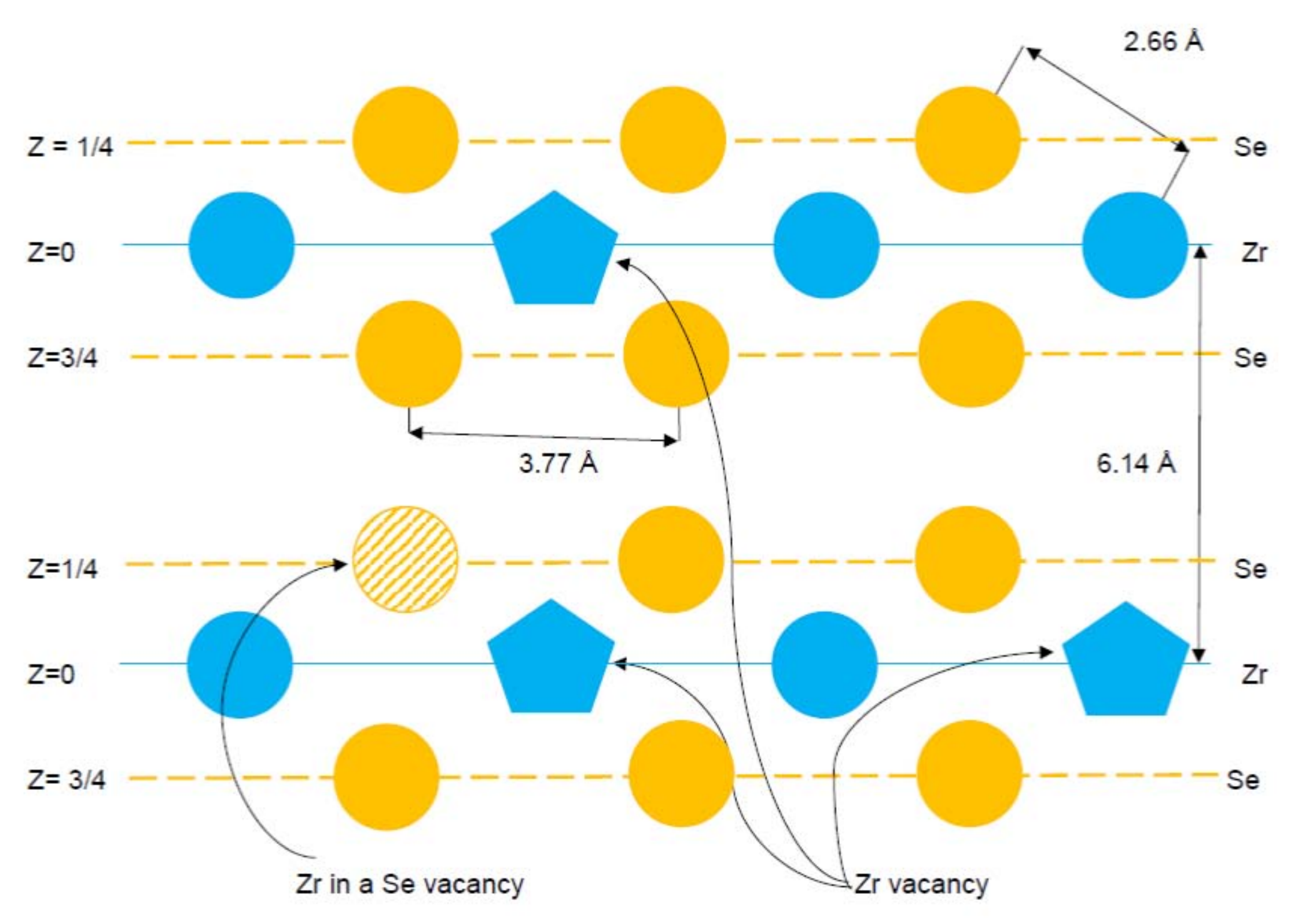}
\caption{(Color on-line) Possible defects or vacancies  configuration in Zr$_x$Se$_2$ with  deficiencies of Zr, according to Gleizes and Jeannin analyzes  \cite {gleit}.Blue circles correspond to Zr atoms,  Se are  in yellow color. Vacancies  can be in different sites and number. }
\end{center}
\end{figure}

ZrSe$_2$ crystalizes in C6-CdI$_2$ type structure. Frequently   may exhibits deviations of stoichiometry, that changes it  physical properties. Many  of those studies were  performed many  years ago by  Van Arkel, MacTaggart and Wadsley, Han and Ness, Gleitzes and Jeannin \cite{van, Mac, Han,  gleit}. The crystalline structure  of  ZrSe$_2$ has a chain-like structure formed by a sequence of stacking atoms in the unit cell, In Fig 1 we show the structure of this compound with  details related to vacancies and  stacking of the structure. As mentioned by these  authors the variant of stacking atoms in C6 type compounds may presents vacancies of Se or Zr depending of the Se/Zr ratio, the crystal structure of the compound has parallel layers consisting of two hexagonal chalcogenes planes  with  one hexagonal Zr plane. The intralayer bonds are strong, whereas the others are quite weak. The compound is highly anisotropic as  seen in  its  physical properties, i.e.  electrical and thermal conductivity, as many others properties  \cite{iso}.
In fact it is important to mention  that defects   in ZrSe$_2$, play an important role in the compound,  as Gleitzes and Jeannin found \cite{gleit}.  They observed  that two type of defects or vacancies  can occur  depending of the Se/Zr ratio.  One important change we found is related to the   semiconducting behavior  to superconducting character. These new characteristics reported here  are  the main result of our investigations. Thus we found that decreasing the Zr content up to some small level ZrSe$_2$   becomes a type two superconductor.

\section{Experimental details}

Synthesis of the ZrSe$_2$ was carried out by solid state reactions.  Zr and Se reactives were   alpha Aldrich  with  purities  of  99.9 \%, and 99.999 \% respectively. The powders were mixed in  an agatha mortar  inside a globe box. The powders with appropriated stoichiometry were sealed  in a quartz tube under a clean atmosphere free of oxygen using first, an argon atmosphere, and evacuated. The synthesize was performed at  a temperature  of 800 Celsius  in different periods of time. In  general we noted that  5 or 6 days of heating is enough  to obtain  a pure compound. Several samples with different stoichiometries, Zr$_x$Se$_2$ were produced. The amount of  Zr content was changed  from x= 0.70 to 0.97 according to the formula Zr$_x$Se$_2$. In this two limits the resulting compound is semiconducting, whereas in between the material becomes superconducting with  small variation in the transition temperature.

Chemical analysis for all  samples,  was performed by Rutherford Backscattering Spectrometry (RBS)   performed using a Pelletron NEC accelerator (National Electrostatics Corp, Middleton, WI) with 3 MeV beam of alpha particles. RBS spectra were collected with a detector at  an angle of 12 degrees  respect to the incident beam. The sample surface was normal to the incident beam. Experimental chemical analysis was fitted to  a theoretically calculated curve by means of the program XRUMP.

\section{X-Ray Diffraction Analysis and Structural Characterization}

Powders of each  sample for X-Ray determination  were  measured at room temperature in a Bruker D8 Advance diffractometer (Bruker AXS GmbH, Karlsruhe, Germany; CuK$_{\alpha1}$ radiation, $\lambda = 1.5405$  $\AA$ and goniometer with a lynx-eye detector (Bruker AXS GmbH). Data were collected from  2$\theta$,  6 - 90 degrees, with 30 kV and 40 mA in the X-ray generator. Figure 2 displays  X-Ray data and analyses by Rietveld. The structural information for each one of the identified phases was obtained from the Inorganic Crystal Structure Database (ICSD) databank \cite{crystal}. Cell parameters, crystal symmetry, and atomic coordinates were introduced in the Rietveld program GSAS \cite{larson} using the graphical interphase EXPGUI \cite{toby}. A modified pseudo-Voigt function was chosen to generate the peak shapes of the diffraction reflections. The refined parameters were zero point and scale factors, cell parameters, half-width, preferred orientation and atomic occupancy factors. Two types of defects proposed by Gleizes \& Jeannin \cite{gleit}] for the ZrSe$_2$ structure were considered in the Rietveld refinements.

\section{Results and discusion}

Single phase of ZrSe$_2$ was found  in all samples with  exception of the sample with  high  superconducting fraction about  10.3\%, where selenium appears as secondary phase. This is shown in first column of Table.  In the crystallographic data for ZrSe$_2$ reported by Van Arkel \cite{van}, Zr is at the origin of the unit cell, while Se occupies the 2d position at (1/3, 2/3, 1/4) forming a layered structure, details are clearly shown  in Fig 1. Each layer has a sandwich-like structure in which Zr atoms form a two-dimensional hexagonal close packing plane and six Se atoms octahedrally coordinate each of them (Fig 1b).  
Of the two types of defects proposed by Gleizes \& Jeannin \cite{gleit}, only one   (that  refers to vacancies of Zr atoms,  located at the origin of the unit cell) leds to good results in our  Rietveld refinements, however as mentioned by Ikari \cite{ikari} Zr also may be found in interticial site of the crystal structure. The second type of defect has to do with   substitution of Se by Zr in the (1/3, 2/3, 1/4) position; since in the refinements the occupancy factors for Se and Zr gave practically 1 and 0 respectively, this defect was  considered absent in the samples.  With this dominant defect we  found that the lattice parameters vary from 3.7722 to 3.7576 $\AA$ for $a$, and from 6.1297 to 6.135 $\AA$ for $c$, as the Se/Zr ratio increases from 2.06 to 2.47 and the superconducting fraction increases from 0\% to 10.3\% as shown in (Table).

\begin{table*}[t]
\caption{ Table shows the characteristics of the studied samples. Samples S1, S2, and S3 are superconducting,   sample S4 is not. Table shows the detected phases, the ICSD code, space group and cell parameters, only in S1 we see a drastic reduction of volume. The details were determined by Rietveld analyses. However the most important change related to superconductivity  was the proportion of Zr\/Se ratio. The main phase was modeled with  the structural type of ZrSe$_2$, code in the ICSD database listed in theTable. Those number were **Calculated from the quantification of identified phases by Rietveld refinement, and ***Calculated from RBS results.}
\resizebox{\textwidth}{!}{%
\begin{tabular}{|l|c|c|c|c|c|}
\hline
Samples                                                                 & \multicolumn{2}{c|}{S1}                                                                                                                   & S2                                                                    & S3                                                                    & S4                                                                    \\ \hline
\begin{tabular}[c]{@{}l@{}}Superconducting\\ fraction (\%)\end{tabular} & \multicolumn{2}{c|}{10.3}                                                                                                                 & 1.3                                                                   & 0.7                                                                   & 0.0                                                                   \\ \hline
Crystal Phases                                                          & Zr$_{0.81}$Se$_2$                                                            & Se                                                                 & Zr$_{0.93}$Se$_2$                                                             & Zr$_{0.93}$Se$_2$                                                             & Zr$_{0.97}$Se$_2$                                                             \\ \hline
ICSD code*                                                              & 652244                                                               & 53801                                                              & 652244                                                                & 652244                                                                & 652244                                                                \\ \hline
Weight (\%)                                                             & 99.76(5)                                                             & 0.24(4)                                                            & 100                                                                   & 100                                                                   & 100                                                                   \\ \hline
Space group                                                             & P-3m1                                                                & P3$_1$21                                                              & P-3m1                                                                 & P-3m1                                                                 & P-3m1                                                                 \\ \hline
Cell parameters ($\AA$)                                                     & \begin{tabular}[c]{@{}c@{}}a = 3.7576(6)\\ c = 6.135(1)\end{tabular} & \begin{tabular}[c]{@{}c@{}}a = 4.40(1)\\ c = 4.948(9)\end{tabular} & \begin{tabular}[c]{@{}c@{}}a = 3.7695(1)\\ c = 6.1437(2)\end{tabular} & \begin{tabular}[c]{@{}c@{}}a = 3.7701(2)\\ c = 6.1354(4)\end{tabular} & \begin{tabular}[c]{@{}c@{}}a = 3.7722(3)\\ c = 6.1297(4)\end{tabular} \\ \hline
Volume ( $\AA$3)                                                            & 75.01(3)                                                             & 83.0(4)                                                            & 75.60(1)                                                              & 75.52(1)                                                              & 75.54(1)                                                              \\ \hline
Se/Zr ratio**                                                           & \multicolumn{2}{c|}{2.47}                                                                                                                 & 2.15                                                                  & 2.15                                                                  & 2.06                                                                  \\ \hline
Zr/Se ratio***                                                          & \multicolumn{2}{c|}{2.44}                                                                                                                 & 1.54                                                                  & 2.38                                                                  & 1.54                                                                  \\ \hline
R$_{wp}$                                                                     & \multicolumn{2}{c|}{0.038}                                                                                                                & 0.064                                                                 & 0.113                                                                 & 0.109                                                                 \\ \hline
R$_p$                                                                      & \multicolumn{2}{c|}{0.028}                                                                                                                & 0.043                                                                 & 0.082                                                                 & 0.074                                                                 \\ \hline
\end{tabular}}

\end{table*}

The crystal structure was determined by X-Ray Diffraction Analysis and the Structural Characterization of the  powders of each sample  at room temperature with  a Bruker D8 Advance diffractometer (Bruker AXS GmbH, Karlsruhe, Germany; using  Cu$K_{\alpha1}$ radiation, $\lambda = 1.5405  \AA$ and goniometer with a lynx-eye detector (Bruker AXS GmbH). Data were collected, in 2$\theta$ from  6 to 90 degrees, with 30 kV and 40 mA in the X-ray generator.  A glass sample holder  was used to  perform the characterization of powders. The X-Ray analysis and structural characterization information   for each sample was  obtained and identified  using  the ICSD databank (see Table). Cell parameters, crystal symmetry, and atomic coordinates were introduced to fit the crystal structure using  Rietveld program. It is interestig to mention that cell parameters determined for this compound coincides with calculation of other authors, see for instances Hussain, et, al \cite{hussain}. A modified pseudo-Voigt function was chosen to generate the peak shapes of  reflections. The refined parameters were zero point and scale factors, cell parameters, half-width, atomic coordinates, isotropic thermal coefficients for each phase.  For the case of the ZrSe$_2$ vacancies there were considered  in the Zr sites and  the occupancy of this site was refined.

\section{Superconducting characteristics}
Figures 3 and 4 show the superconducting behavior of the compounds.  Fig. 3 displays four curves of Magnetization - Temperature of four different samples with different Zr vacancies,  the four samples show small variation of the transition critical temperature but clearly defined from 8.14, 8.31, 8.34, and 8.59 onsets in Kelvins. Those curves were determined in ZFC an FC in order to determine  the fraction of superconductivity. In Fig. 4 we present the behavior of the  two critical  fields. 

As  before mentioned,   one importan aspect on this study is related to  defects configuration  in  the  crystalline structure of  the compound.  The   defects  on  the (100) plane,  as found by Gleizes and Jeannin, \cite{ gleit} is quite important because some of the physical properties are depending on it, for instance in   the Se/Zr ratio. The number  of defects  (vacancies) is a critical aspects for the physics behavior,  the number of defects, (vacancies)  is important and turn to be very  complicated in elucidation of the involved physics. The electronic properties  depend of the missing  atom and position on the cell.  The superconducting characteristics also are  depending on the number of  vacancies.  According to this,  the compound, could be  semiconductor, metal, or  superconductor.

Onuki et al, \cite{onuki} and Thompson \cite{tom} have studied the electrical properties of TiS$_2$ and ZrSe$_2$, they  found  that in   T${^2}$ the  resistivity behavior,   above 50 K, behaves as a degenerated semiconductor similarly as occurs in  ZrSe$_2$. This  is a  consequence that the number of defects  change the size of the band gap.  Our Fig. 5 shows a similar picture of the distribution of  vacancies in the (100) plane, considering only  Zr atoms as our compositions; with  Zr vacancies about 0.75, 0.80, and 0.85, this figure is quite similar to in the Gleizes and Jeannin paper. It is worth mentioning that in  Onuki et al, \cite{onuki} they mention similar deficiencies in the  TiS$_2$. The  changes on the density of states is depending on  the number of missing Ti atoms, and therefore  physical properties displays a change from a weak metallic behavior to  a semiconductor  and to a metal. In our study  Zr deficiencies provoke  a change into the  t$_{2g}$ band. This is our main consideration to the change from  semiconducting to  a superconductor. It is important to mention that in this compound six valence
bands are primarily derived from Se 4$p$ orbitals while the conduction bands are derived  from the Zr 4$d$ orbitals\cite{brauer}. 
It is worth mentioning the importance of the band filling in the compound:  the six valence bands of this compound  are primarily derived from chalcogen p orbitals while the conduction bands are derived from the transition metal $d$ orbitals. The valence band can hold 12 electrons per unit cell, so after filling the valence band, no $d$ electrons remain for TiS$_2$ and ZrSe$_2$, making them semiconductors with indirect band gaps, Eg ~ 0.2 eV and Eg ~ 1 eV respectively. Friend et al \cite{friend} found that TiS$_2$ is a degenerate semiconductor which carriers do not arise from $p$, $d$ band overlap, but from partial occupation of t$_2g$ band resulting by Ti excess. and the compound becomes a degenerated semiconductor  and the high conductivity can be attributed to charge transfer from self intercalated Ti atoms in the Van der Waals inter layer space.

as mentioned by Brauer, et. al. \{brauer, and Onuki et al \cite{onuki}, they investigate  an interesting behavior related to the band filling in  TiS$_2$.    The band structure of   the valence band  can hold 12 electrons per unit cell, and no electrons remains after filling, thus TiS$_2$ and ZrSe$_2$ becomes semiconducting, with  a band gap about 1 to 1.3 eV. The six valence bands are primarily derived from chalcogen $p$ orbitals while the conduction bands are derived from the transition metal $d$ orbitals. The valence band can hold 12 electrons per unit cell, so after filling the valence band, no $d$ electrons remain for TiS$_2$ and ZrSe$_2$, making them
semiconductors with indirect band gaps, Eg, Once the band is not completed filled with the Zr vacancies, then superconductivity characteristic arises. In other  situations the compound behaves as  a semiconductor. 

 It is important to mention that the superconducting characteristics  show only a small diamagnetic fraction, so  the superconductivity fraction is very small, or  we can call  filamentary.  In Fig. 2 we show those measurements,  in  plots of Magnetization - Temperature, (M-T) determined in zero field cooling (ZFC) and field cooling (FC) at  10 Oe. Fig 3 shows the critical fields, top figure shows H$C_1$ fitted to a parabolic function with critical temperature of T$_C$ = 8.8 K. Bottom plot displays H$C_2$. fitted only to a linear function. The magnitude of the two critical fields is of the order of 170,  and 2500 Oe for the two fields. The superconductor is a  type two,  with  thermodynamic field about H$C$ = $650$ Oe, so the the Ginzburg-Landau parameter,  $\kappa$ is 2.70, is in the strong coupling limit \cite{kittel}.
In order to have more insight into the characteristics of this new superconductor, various samples were used to measure the specific heat in function of temperature,  close to the superconducting temperature, we never see the specific heat jump related to the  transition.This is clearly indicative that the superconductivity  is only a very small fraction, and therefore  filamentary.  Resitivity mneaurements were not measured, because the compound was powder, and almost imposible to form a bulk compact material.

\section{Conclusions}

 We have  found and studied  a new  superconducting compound with composition ZrSe$_2$, which presents a transition temperature about 8.14 - 8.59 K with   Zr vacancies from about 0.75, 0.8 and 0.85, 
out of this small window range  the semiconducting compound is as already determined.

\section*{Acknowledgments}
We  thank to DGAPA UNAM  project IT106014,  and to A. L\'opez-Vivas, and A. Pompa-Garc\'ia (IIM-UNAM), for help in technical problems.

\end{document}